\documentclass[runningheads]{llncs}
\usepackage[caption=false]{subfig}
\usepackage{amssymb}
\usepackage{amsmath}
\usepackage{stfloats}
\usepackage{multirow}
\usepackage{subfig}
\usepackage{makecell}
\usepackage{graphicx}
\usepackage[misc]{ifsym}
\usepackage[hyphens]{url}
\usepackage[hidelinks]{hyperref}
\usepackage{tikz}
\usepackage{adjustbox}
\usepackage{booktabs}
\usepackage{rotating}
\usepackage{multirow}
\usepackage{hyperref}
\usepackage{tabularx,ragged2e}
\usepackage{blindtext}
\usepackage{pifont}
\newcommand{\keywords}[1]{\par\addvspace\baselineskip
\noindent\keywordname\enspace\ignorespaces#1}
\usepackage{svg}
\newcommand{\orcid}[1]{\href{https://orcid.org/#1}{\includesvg[width=10pt]{orcid}}}
\newcommand{\blue}[1]{\textcolor{blue}{{#1}}}

\begin{document}


\title{Detecting High-Quality GAN-Generated Face Images using Neural Networks}


%
%

\author{Ehsan~Nowroozi \inst{1,2}\textsuperscript{(\Letter)} Mauro~Conti \inst{2}, and Yassine~Mekdad\inst{3}}
\authorrunning{E. Nowroozi et al.}

\institute{Faculty of Engineering and Natural Sciences (FENS), Center of Excellence in Data Analytics (VERİM), Sabanci University, Istanbul Turkey 34956.\\
\email{ehsan.nowroozi@sabanciuniv.edu}\\
\and
Department of Mathematics, University of Padua, 35121, Padua, Italy,
\email{\{nowroozi,conti\}@math.unipd.it}
\and
Cyber-Physical Systems Security Lab, Department of Electrical and Computer Engineering, Florida International University, Miami, FL 33174 USA\\
\email{ymekdad@fiu.edu}}

\maketitle
%
%

\begin{abstract}
In the past decades, the excessive use of the last-generation GAN (Generative Adversarial Networks) models in computer vision has enabled the creation of artificial face images that are visually indistinguishable from genuine ones. These images are particularly used in adversarial settings to create fake social media accounts and other fake online profiles. Such malicious activities can negatively impact the trustworthiness of users' identities. On the other hand, the recent development of GAN models may create high-quality face images without evidence of spatial artifacts. Therefore, reassembling uniform color channel correlations is a challenging research problem. To face these challenges, we need to develop efficient tools able to differentiate between fake and authentic face images.

In this chapter, we propose a new strategy to differentiate GAN-generated images from authentic images by leveraging spectral band discrepancies, focusing on artificial face image synthesis. In particular, we enable the digital preservation of face images using the Cross-band co-occurrence matrix and spatial co-occurrence matrix. Then, we implement these techniques and feed them to a Convolutional Neural Networks (CNN) architecture to identify the real from artificial faces. Additionally, we show that the performance boost is particularly significant and achieves more than 92\% in different post-processing environments. Finally, we provide several research observations demonstrating that this strategy improves a comparable detection method based only on intra-band spatial co-occurrences. 
\end{abstract}

\section{Introduction}
The recent advancements of the machine and deep learning models created challenging models referred to as Generative Adversarial Networks (GANs) \cite{GoodFelow2014}. These networks enable the creation of totally artificial images from scratch and added extras such as image editing, attribute manipulation, and style transfer. Apparently, non-expert humans may generate realistic fake photos because of the remarkable efficiency of deep neural networks. In particular, the most newer versions of GAN networks can produce very high-quality pictures, such as facial images, which may effectively fool the human user \cite{Brock2018}.
Given several real-world applications, the potential misuse of GAN synthetic content poses a significant concern, thus necessitating the creation of image forensics systems capable of distinguishing between genuine and artificial images.
Different approaches have been suggested in multimedia forensics to identify whether an image is real or GAN-generated. The most modern approaches are focused on CNN models and can achieve outstanding results.
In \cite{Nataraj2019}, the authors proposed an approach that can produce extremely excellent detection performance of GAN-generated images. The approach generates the co-occurrence matrix from different color image bands, then feeds them into the Convolutional Neural Network (CNN). 

By observing that recent GAN models may create extremely high-quality images with virtually imperceptible spatial errors (if any), reconstructing a consistent relationship between colors is likely more challenging. In this chapter, we enhance the recognition of GAN-produced images by expanding the techniques presented in \cite{Nataraj2019}. More specifically, we supply the CNN detection method using cross-band and gray-level co-occurrences that are estimated individually on the single-color bands.
Our experimental results rely on the well-known StyleGAN model \cite{StyleGAN-Karras} \cite{StyleGAN2-Karras} that creates higher quality images than prior models such as ProGAN \cite{Prog-Karras}, attempting to make the detection task more difficult. In comparison to the intra-band technique suggested in \cite{Nataraj2019}, our CNN detector achieves almost optimal detection performance, which has far higher robustness to post-processing.

Given the vulnerability of several approaches regarding the recognition of JPEG compressed faces, we re-trained the model using JPEG compressed faces to recognize JPEG-compressed artificial images. 
Our experimental results demonstrated that the JPEG-aware models could perfectly detect with high level of accuracy the JPEG compressed GAN images in matching and mismatching scenarios. In other words, the JPEG quality factors are different when they are used only for training and validation. In this study, our experimental results exhibit the effectiveness of the JPEG-aware models in the post-processing environments performed to the images previously to JPEG compression, thus confirming that the JPEG-aware versions of the detectors are particularly resistant versus post-processing.
\subsection{Organization}
The rest of our chapter is organized as follows: We briefly cover research efforts on GAN image recognition in Section II, emphasizing relevant approaches to color space analysis. In Section III, we present the review and approval techniques. Afterward, we describe our approach in Section IV. Then, we provide and analyze the outcomes of our results in Section V. Finally; we conclude our chapter with some last considerations in Section VI.

\section{State of the Art}
Prior works have suggested several methods to differentiate between artificial images generated through GANs and genuine ones. Some algorithms use particular facial traces \cite{Matern2019}. For instance, the authors in \cite{Yang2019} demonstrated that by evaluating the placements of facial feature spots and using them for training an SVM, we can identify GAN-generated images. Other approaches that utilize color information to disclose GAN features are more similar to the one given in \cite{McCloskey2018}. In this approach, the authors presented two metrics based on comparisons among color information and intensity for the detection by examining GAN networks' behavior and color cues. 

In\cite{Cozzolino214} \cite{Barni2017}, the co-occurrence features, which are frequently calculated on residual images, are commonly employed in detection techniques for identifying or localization modifications. These approaches commonly consider the SPAM features \cite{Pevny2010}, the rich feature model \cite{Fridrich2012}, and the rich features modeling techniques for color images \cite{Goljan2014}, which were first developed for image steganography. 
Additional studies have been suggested by \cite{Li2020}, the authors incorporated the assessment of the color channel and the SPAM-like properties to detect GAN images. In this case, the co-occurrences matrices are retrieved from truncated residuals of numerous color components and a truncated residual image constructed by merging the Red, Green, and Blue bands. Afterward, the co-occurrences matrices are concatenated into a feature set and then used to train the SVM. 

Convolutional Neural Networks (CNN)-based approaches have recently been suggested by \cite{Man2020}, \cite{Marra2019}. These approaches outperform prior techniques based on fundamental machine learning and handcrafted features. In \cite{Marra2019}, the authors presented an incremental learning strategy to alleviate the necessity for large trained data from all the various GAN models. 
In a recent paper \cite{Nataraj2019}, the authors demonstrated that providing as input the co-occurrence matrix (that is performed on the source images) to the CNN model can significantly improve the face GAN identification challenge when compared to co-occurrences extracted features from noise residuals.

In this chapter, we benefit from the increased effectiveness of the CNN model and develop a methodology to detect dissimilarities among the color components by observing the co-occurrence cross-band matrices.
\section{Cross Co-occurrences Feature Computation}
It was recently demonstrated that discrepancies in pixel co-occurrences might reveal GAN content. In \cite{Nataraj2019}, the authors recommended that the three co-occurrence matrices generated on the Red, Green, and Blue channels of the images can be fed into a CNN as a tensor of three matrices. Afterward, the network will learn unique characteristics from the co-occurrence matrices by learning with GAN and real images, thus outperforming state-of-the-art approaches. Accordingly, we refer to \textit{Co-Net}, the CNN model that is trained on co-occurrence matrices.

In this study, we observe that recreating consistent relationships across color bands is a challenging task for GANs. Therefore, to leverage the correlations between color bands, we consider estimating cross-co-occurrences and feeding them into the CNN model simultaneously with the spatial co-occurrences estimated on the single bands. In this case, we denote \textit{Cross-Co-Net} as the name given to the network that has been trained using this strategy. It is worth mentioning that cross-band features should be resistant to the standard post-processing operations that are stressing on spatial pixel correlations instead of cross-band features. 

Given an image $I = (a, b, c)$ of size $H \times V \times 3$, we note Red, Green, and Blue the 2D matrix channels, respectively. For a displacement $\tau = (\tau_a, \tau_b)$, we define the spatial co-occurrence matrix of channel Red, Green, and Blue as follows: \\

\begin{equation}
W_{\tau}(x,y; Red)  =  \sum_{a=1}^H \sum_{b=1}^V 
\begin{cases} 1 & \text{$if$ $\tau(a,b) = x$ $and$  $Red(a + \tau_a,b + \tau_b) =y$}\\ 0 & \text{$otherwise$}
\end{cases}
\end{equation}

\begin{equation}
W_{\tau}(x,y; Green)  = \sum_{a=1}^H \sum_{b=1}^V 
\begin{cases} 1 & \text{$if$ $\tau(a,b) = x$ $and$ $Green(a + \tau_a,b + \tau_b) =y$}\\ 0 & \text{$otherwise$}\end{cases}
\end{equation}

\begin{equation}
W_{\tau}(x,y; Blue)  = \sum_{a=1}^H \sum_{b=1}^V 
\begin{cases} 1 & \text{$if$ $\tau(a,b) = $ $and$ $Blue(a + \tau_a,b + \tau_b) =y$}\\ 0 & \text{$otherwise$}\end{cases}
\end{equation}

For Red, Green, and Blue channels, I(a, b, 1), where $x,y$ are integers between 0 and 255. We note that \textit{RG, GB, and RB} stands for \textit{Red and Green}, \textit{Green and Blue}, and \textit{Red and Blue}, respectively. In what follows, we describe our construction of the cross-co-occurrence matrix (spectral) for the channels Red, Green, and Blue:

\begin{equation}
W_{\tau'}(x,y; RG) = \sum_{a=1}^H \sum_{b=1}^V 
\begin{cases} 1  & \text{$if$ $I(a,b,1) = x$ $and$ $I(a + \tau_a',b + \tau_b',2) = y$}\\ 0 & \text{otherwise}\end{cases}
\end{equation}
\begin{equation}
W_{\tau'}(x,y; GB) = \sum_{a=1}^H \sum_{b=1}^V 
\begin{cases} 1  &  \text{$if$ $I(a,b,1) = x$ $and$ $I(a + \tau_a',b + \tau_b',2) = y$}\\ 0 & \text{otherwise}\end{cases}
\end{equation}
\begin{equation}
W_{\tau'}(x,y; RB) = \sum_{a=1}^H \sum_{b=1}^V 
\begin{cases} 1  & \text{$if$ $I(a,b,1) = x$ $and$ $I(a + \tau_a',b + \tau_b',2) = y$}\\ 0 & \text{otherwise}\end{cases}
\end{equation}
\\\\
Rather than being given intra-channel, the offsets $\tau' = (\tau_a', \tau_b')$ is used inter-channel, — in other words across the bands. The offset in the cross-band scenario accounts for two impacts: the difference band as well as the different spatial position. For cross-co-occurrence analyses, a $\tau'$ of [0, 0] represents the condition where there is not any movement in the image pixel; on the other hand, there is a movement in spectral bands that we considered pixel conditions [1, 1].
The six tensors that comprise the Cross-Co-Net network's ($\digamma_{\tau, \tau'}$) input are the three co-occurrence matrices for the channels [Red, Green, and Blue] and the three cross-co-occurrence matrices for the couple [Red Green, Red Blue, and Green Blue], namely:
\begin{eqnarray}
\digamma_{\tau, \tau'}(x,y; :) = [W_{\tau}(x,y; Red),  W_{\tau}(x,y; Green), \nonumber W_{\tau}(x,y; Blue), \\ W_{\tau'}(x,y; Red \ and \ Green), W_{\tau'}(x,y; Red \ and \ Blue), \\ W_{\tau'}(x,y; Green \ and \ Blue)]. \nonumber
\end{eqnarray}
In Figure 1, we present our proposed CNN detector architecture for high-quality GAN-generated face images. First, we input the source image to the spatial and spectral co-occurrence matrices. Then, we output the corresponding features to the Convolutional Neural Network, which will eventually decide whether the input source image is a real face image or a GAN-Face image.

\begin{figure}[h]
    \label{CNN}
	\centering
	\includegraphics[width=\textwidth]{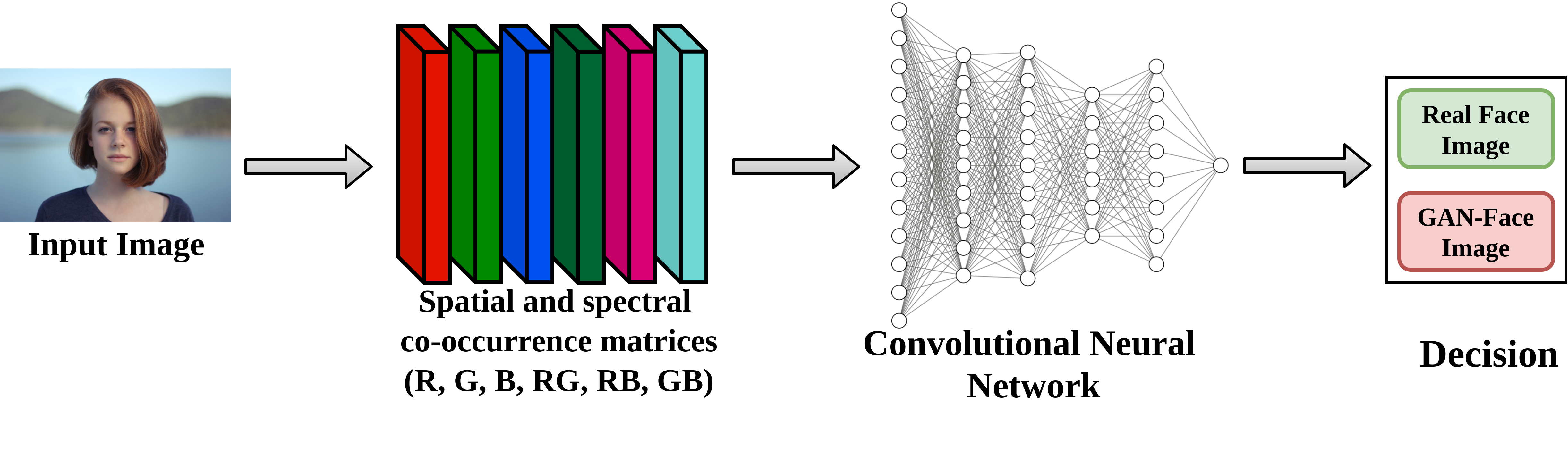}
	\caption{The considered CNN-based detector scheme for high-quality GAN-generated face images.}
\end{figure}

\section{Evaluation Methodology}
In the absence of malicious attackers, we anticipate a good performance from the Cross-Co-Net and Co-Net detectors. Here, we refer to the resiliency of the network. Moreover, we improve the security of the classifiers by re-training them with the attacks that degrade their effectiveness. In this section, we present the experimental procedures to analyze the performance of our approach.

\subsection{Datasets}
We examined large real-world datasets with legitimate and malicious facial images, namely, StyleGan2 \cite{StyleGAN2-Karras} and VIPPrint datasets. Interestingly, both datasets are considered high-quality and challenging datasets.

\subsubsection{StyleGan2 Dataset.}
\begin{figure}%
	\centering
	\subfloat[Authentic FFHQ images]{{\includegraphics[width=6.5cm]{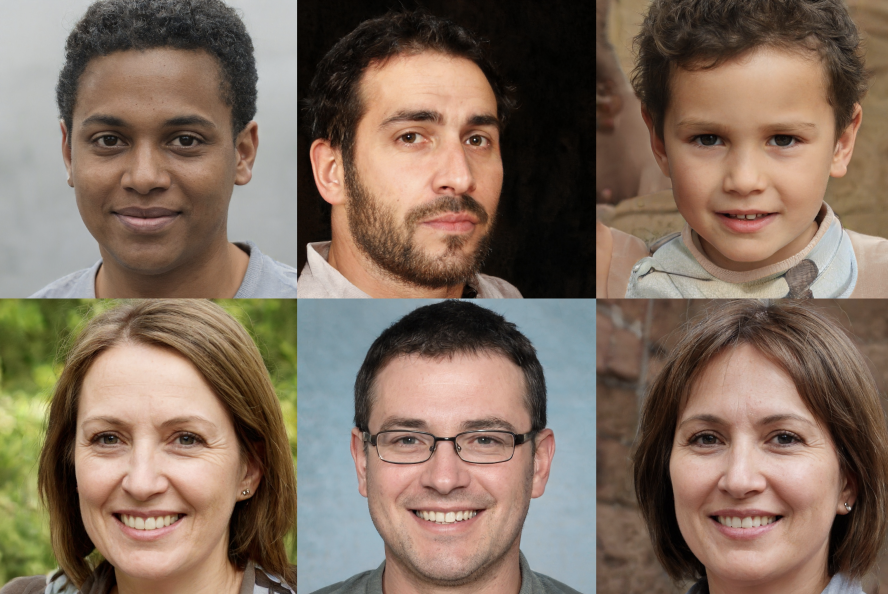} }}%
	\subfloat[StyleGAN2-generated images]{{\includegraphics[width=5.5cm]{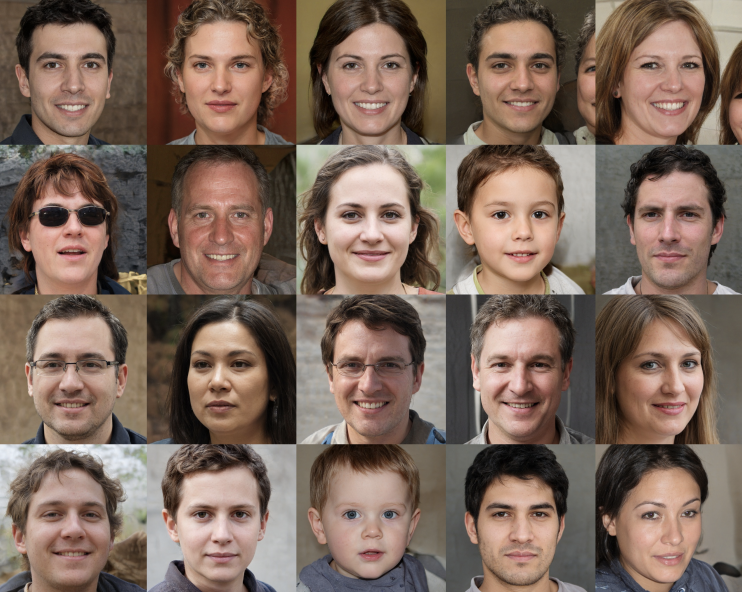} }}%
	\caption{\blue{Illustrations of natural FFHQ and StyleGAN2-generated images that are hardly distinguishable}}%
	\label{fig:example1}%
\end{figure}

The StyleGAN2 image dataset \cite{StyleGAN2-Karras} has recently been suggested as a refinement to the original StyleGAN design \cite{StyleGAN-Karras} and has the capability of producing outstanding results, and thus by generating exceedingly high-quality artificial images. Additionally, face images developed using StyleGAN2 are hardly distinguishable from authentic faces (without visual effects on the face and backgrounds area), making their identification extremely difficult.\\
To train the StyleGAN dataset, we use the Flicker-Faces-HQ (FFHQ) database. It consists of a set of 70000 high-quality images of human faces created as a standard for GAN networks. These images are provided in PNG image format with a size of $1024\times1024$, and illustrate a wide range of ages, genders, races, and so forth. The dataset, which includes faces with accessories like eyeglasses, sunglasses, hats, and so on, is automatically cropped and aligned. Figure \ref{fig:example1} illustrates different natural and generated StyleGan2 facial images.
\subsubsection{VIPPrint Dataset.}
\begin{figure}%
	\centering
	\subfloat[FFHQ]{{\includegraphics[width=4cm]{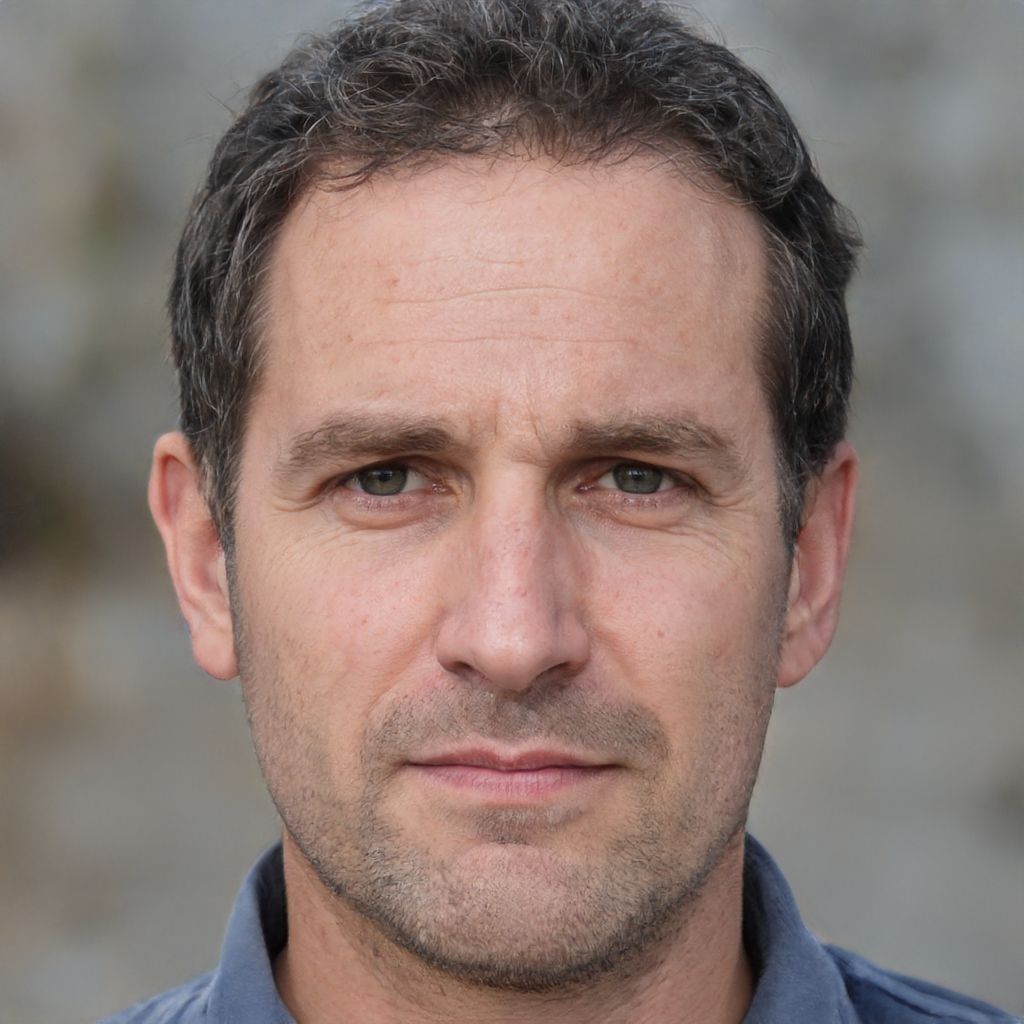} }}%
	\subfloat[FFHQ (Printed-Scanned)]{{\includegraphics[width=4cm]{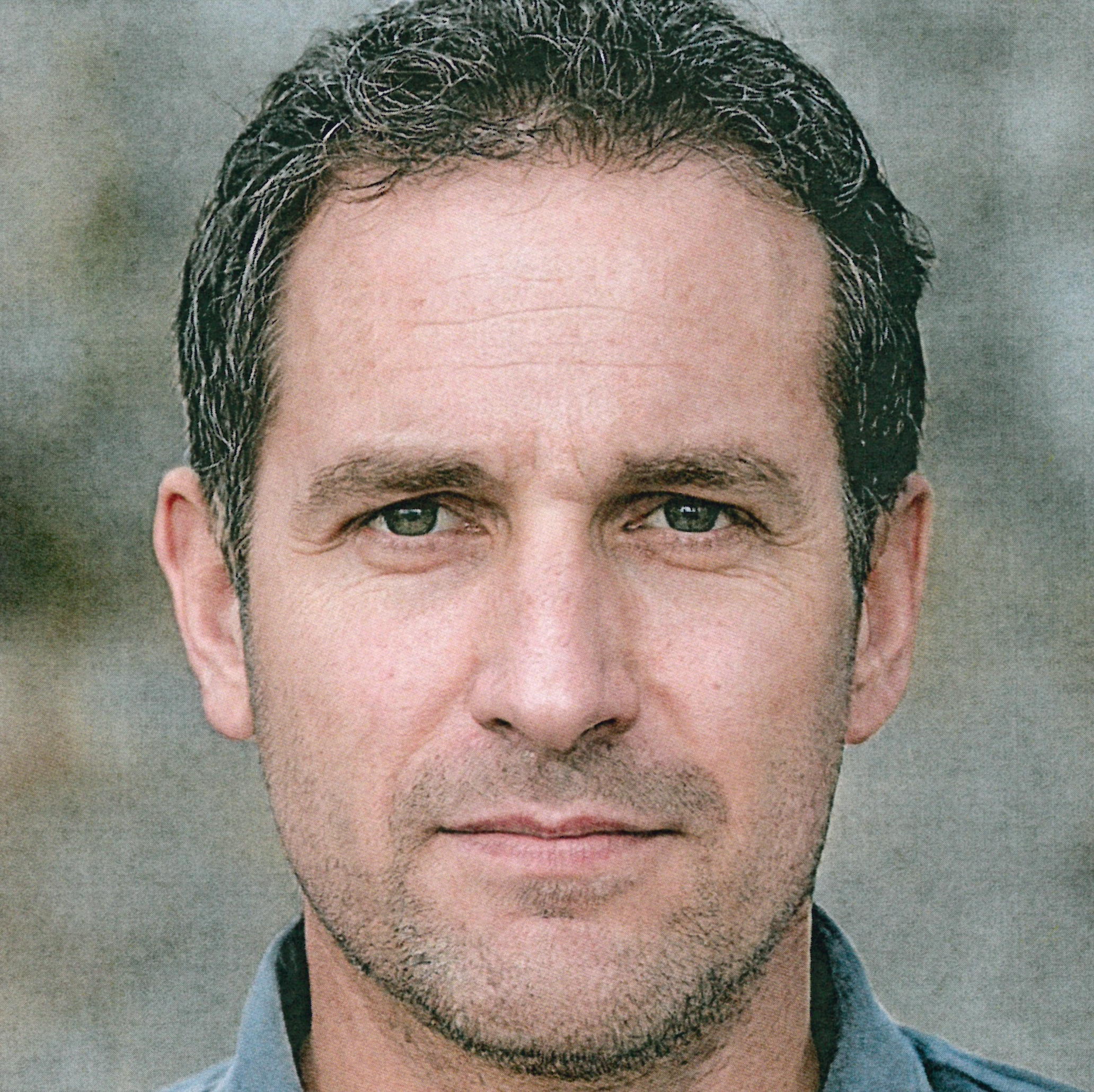} }}%
	\qquad
	\subfloat[StyleGAN2]{{\includegraphics[width=4cm]{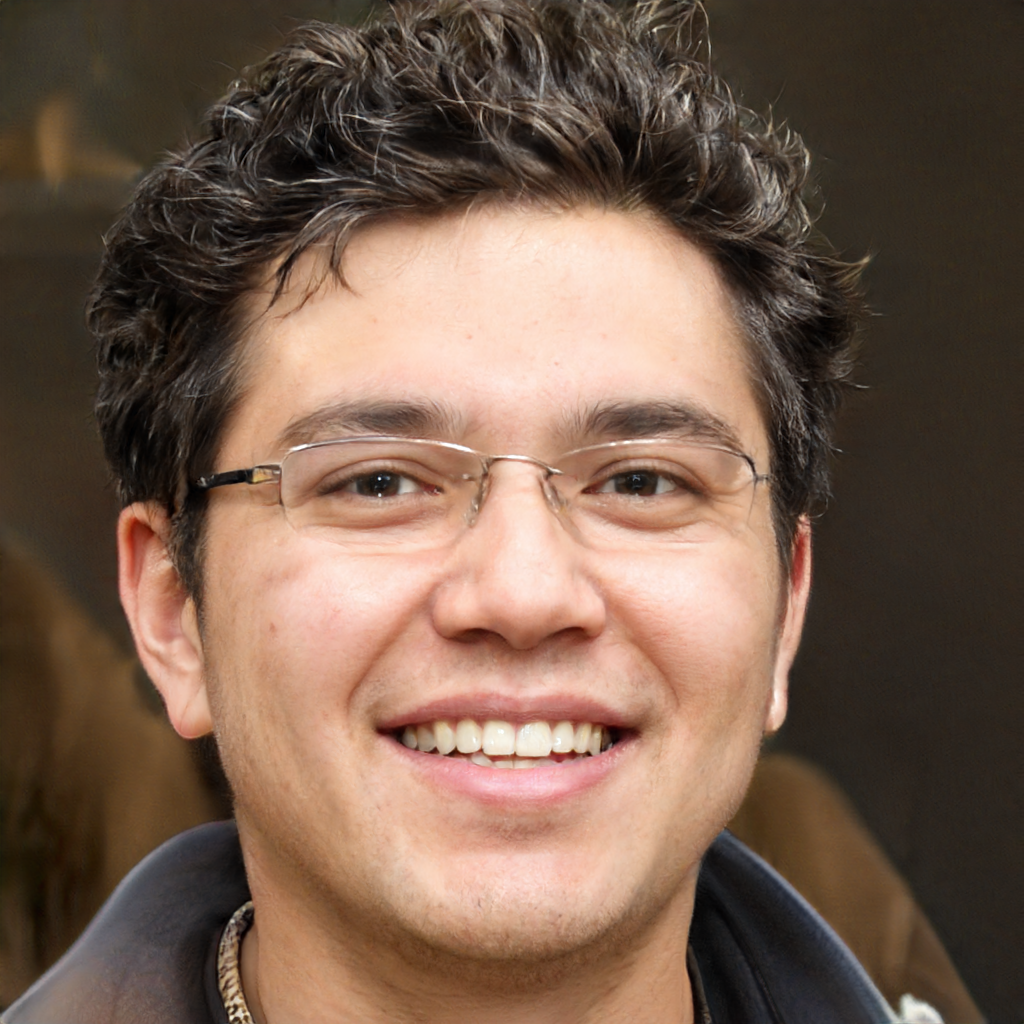} }}%
	\subfloat[StyleGAN2 (Printed-Scanned)]{{\includegraphics[width=4cm]{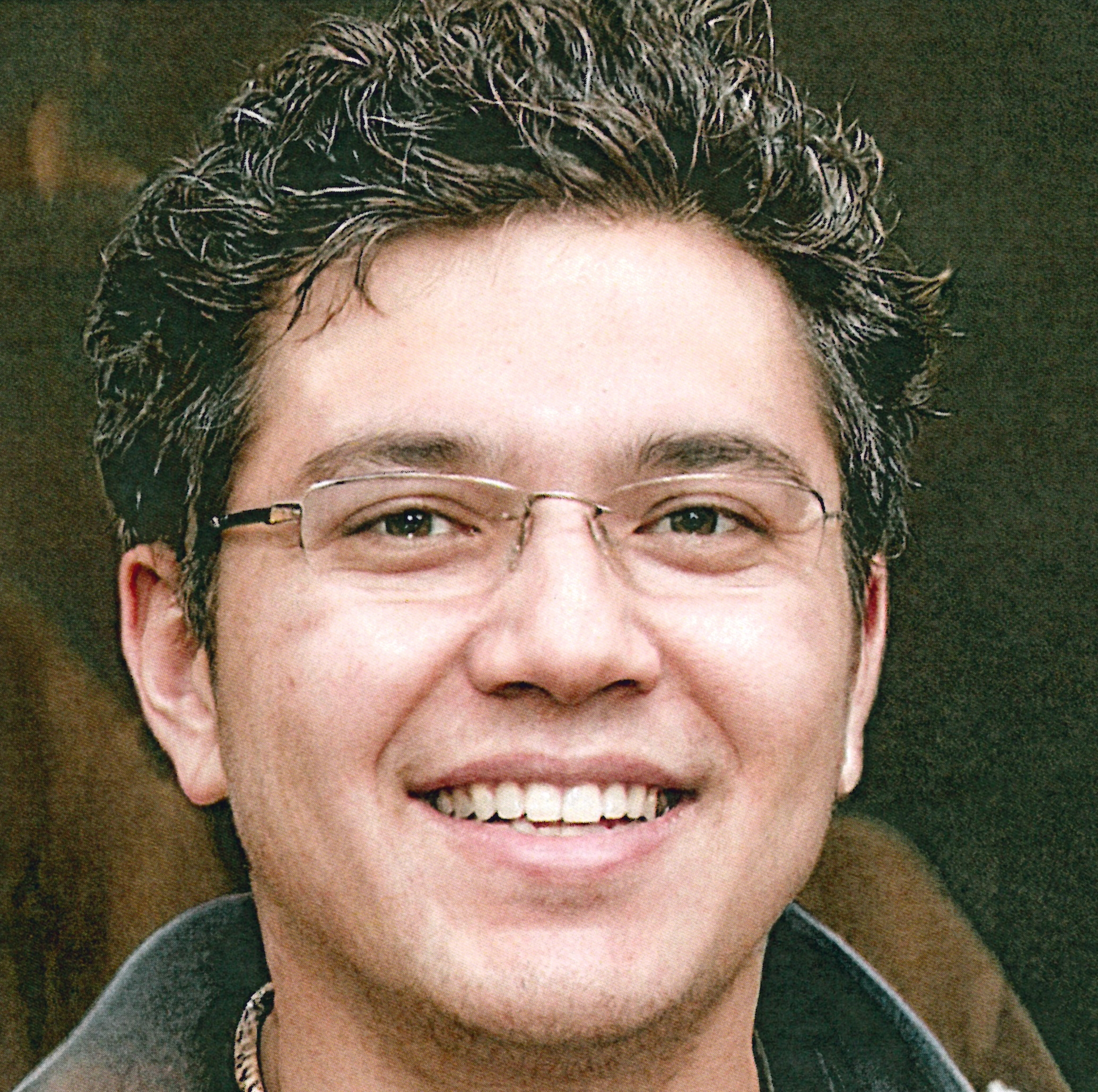} }}%
	\caption{\blue{Illustrations of authentic FFHQ and StyleGAN2 generated face images with their printed-scanned versions.}}%
	\label{fig:example2}%
\end{figure}

VIPPrint image dataset \cite{Ferreira2021} is primarily focused on evaluating artificial image identification and source linkage algorithms on a one-of-a-kind large dataset of printouts. VIPPrint's research is mostly focused on recognizing a particular printer's identity. Each printer incorporates one-of-a-kind effects into its printed colored materials. The VIPPrint approach recognizes these impacts and employs them in developing a validation system. The primary set of face image examples from the FFHQ dataset was considered to generate VIPPrint images \cite{StyleGAN-Karras}. In this dataset, a total number of 40000 Kyocera TaskAlfa3551ci face images have been printed and scanned. Figure \ref{fig:example2} illustrates the source and generated StyleGan2 digital, printed, and scanned versions. 

\subsection{Network Architecture}
In the network architecture, we employed the same network structure as in \cite{Nataraj2019}, which was previously utilized to recognize GAN images. In particular, we used six convolutional layers followed by a single fully-connected layer in the network (we refer to as Cross-Co-Net). The only difference we performed was the first input because we had a six bands input rather than three in our scenario (we refer to as Co-Net). In Figure 4, we illustrate the pipeline of our network architecture, and we describe its structure as follows\footnote{For each convolution layer; the stride is fixed to one.}: \\
\begin{itemize}
\item The first convolutional layer has 32 filters of size $3 \times 3$, then followed by a ReLu layer.
\item The second convolutional layer has 32 filters of size $5 \times 5$, then followed by a max pooling layer.
\item The third  convolutional layer has 64 filters of size $3 \times 3$, then followed by a second ReLu layer.
\item The fourth convolutional layer has 64 filters of size $5 \times 5$, then followed by a second max pooling layer.
\item The fifth convolutional layer has 128 filters of size $3 \times 3$, then followed by a third ReLu layer.
\item The sixth convolutional layer has 128 filters of size $5 \times 5$, then followed by a third max pooling layer.
\item Finally, a dense layer with 256 nodes and followed by a sigmoid layer.
\end{itemize}
\begin{figure}[h]
    \label{CNN}
	\centering
	\includegraphics[width=\textwidth]{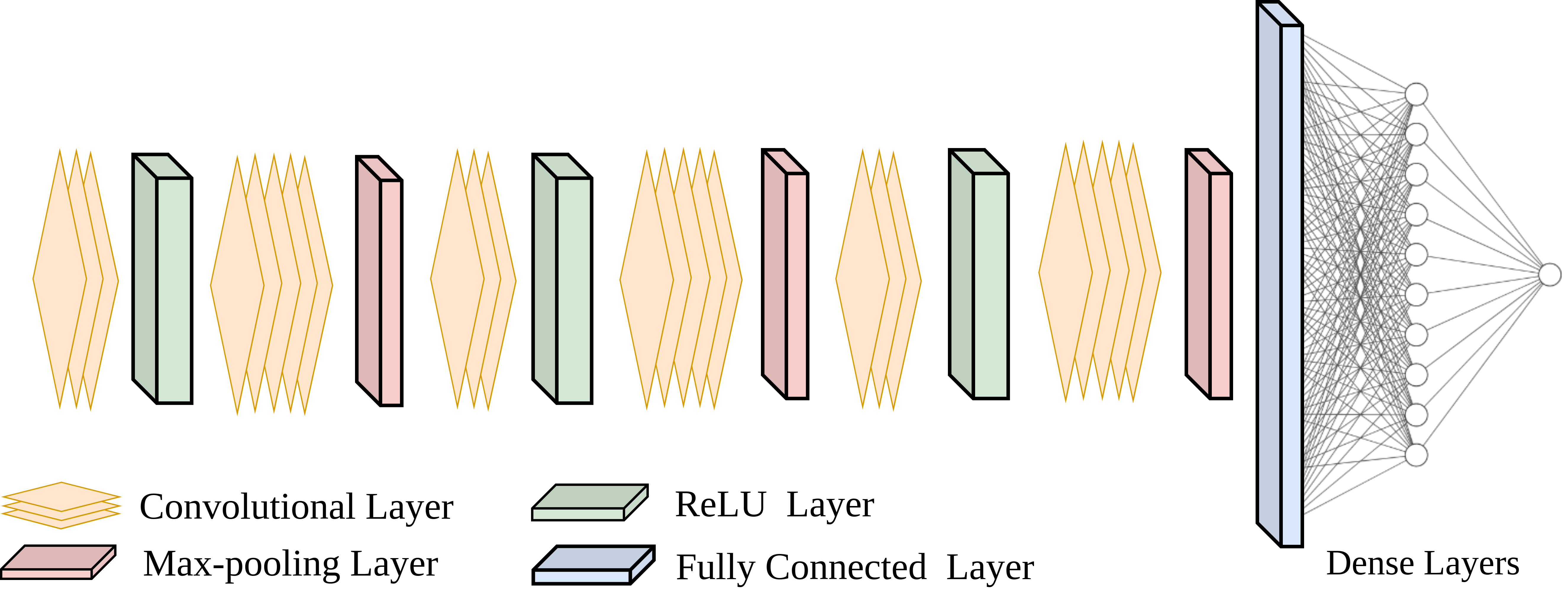}
	\caption{Pipeline of the proposed network architecture.}
\end{figure}

\subsection{Resilience Analysis}
We evaluated the effectiveness of resiliency against several post-processing techniques. In particular, we compared the Cross-Co-Net with the network proposed in \cite{Nataraj2019}. Later, we deployed geometric manipulations such as resizing, rotating, zooming, cropping, and filtering operations (e.g., median filtering, blurring, and contrast enhancements such as gamma correction and adaptive histogram equalization (AHE)). In terms of resizing, we downscaled the images to [0.9, 0.8, and 0.5] as scaling factors and upscaled (referred to as zooming) the utilizing factors [1.1, 1.2, and 1.9], with bicubic interpolation. For rotational operations, we employed bicubic interpolation to assess [5, 10, and 45] angles. We performed the cropping by considering $880 \times 880$. For the filtering, we chose the window sizes for median filtering and blurring at [$3 \times 3$, and $5 \times 5$] respectively. We additionally ran a test versus Gaussian noise in a real-world scenario applying different standard deviations [0.5, 0.8, and 2] and zero means. We considered additional operations to assess the model's robustness, including gamma correction with parameters [0.8, 0.9, and 1.2] and adaptive histogram equalization with a clip parameter of 1.0. Furthermore, we evaluated the robustness of our model against two techniques that follow after each other, including blurring with window size $3 \times 3$ and a kernel  [[-1, -1, -1], [-1, 9, -1], [-1, -1, -1]] for sharpening operation. \\
Our analysis demonstrates that the JPEG-aware version verified the effectiveness of the Cross-Co-Net training technique. It is noteworthy since the attackers often use JPEG compression in digital image forensics to eliminate modification traces. Additionally, we evaluated the effectiveness of the JPEG-aware Cross-Co-Net detector by considering the compression post-processing procedures. For the cases of resizing with scaling factor 0.9, median filtering with a window size of 3 3, and Gaussian noise with a standard deviation of 2, we created a JPEG compressed version of the processed images, along with the compression performed with various Quality Factors (QFs). 

\section{Experimental Results}
In this section, we present the experimental results obtained to evaluate the effectiveness of our newly suggested technique using two different challenging datasets, StyleGAN2 and VIPPrint.

\subsection{Experimental Settings}
For the experiments, we used 20000 authentic FFHQ and 20000 GAN-generated images from the StyleGAN2 dataset, divided as follows for both authentic and GANs: We considered 12000 datasets for training, 4000 for validation, and 4000 for testing. Regarding the VIPPrint dataset, we divided 40000 images for both authentics and GANs as follows: We considered 20000 print-scanned images for training, followed by 10000 for validation and 10000 for testing.

Additionally, we used stochastic gradient descent (SGD) as the optimizer, with a learning rate of 0.01, the momentum of 0.9, the batch size of 40, and training epochs of 40. We implemented our network using TensorFlow's Keras API for training and testing. We also trained the Co-Net network with the same settings to guarantee a fair evaluation.
We used the OpenCV package in Python to post-processing the robustness of our experiments. We considered a total of 2000 per-class images from the test set for each processing operation and parameter.
To identify GAN-generated face images, we constructed the JPEG-aware Cross-CoNet model and utilized the following quality factors [75, 80, 85, 90, and 95]. With 15000 images, we selected 3000 images for training for each quality factor and 5000 images for validation and testing in the real and GAN classes. Then, we retrained the model over 40 epochs, with SGD as an optimizer. We used a similar set of test images to evaluate the robustness of the JPEG-aware model against post-processing.
\subsection{Performance and Robustness of The Detector}
In this subsection, we evaluate the performance of Cross-Co-Net and Co-Net detectors and their robustness in the existence of different post-processing. 

\subsubsection{Cross-Co-Net and Co-Net Performance.}
For the StyleGAN2 detection task from a test set, we remark in Table \ref{tab0} that Cross-Co-Net achieved a test accuracy of 99.80 percent, which is slightly higher than Co-Net of 98.25 percent in the unaware scenario. Additionally, 99.53 percent test accuracy was obtained for Cross-Co-Net and 98.60 percent for Co-Net for the VIPPrint dataset. Overall, the Cross-Co-Net has a significant advantage over Co-Net in post-processing robustness. For the JPEG-aware scenario, the Cross-Co-Net achieves an average accuracy of 94.40 percent on the styleGAN2 datatest and an accuracy rate of 93.80 percent by retraining JPEG-aware versions of the Co-Net model, with a small improvement of Cross-Co-Net over Co-Net that was similar. The JPEG-aware Cross-Co-Net for a VIPPrint dataset acquires an average accuracy of 93.05 percent on the test set for JPEG genuine and fake faces with similar quality characteristics. For the JPEG-aware of Co-Net using a VIPPrint dataset the training resulted in test accuracy of 92.56 percent.

\begin{table}[h!]
\centering
	\caption{Accuracies of Cros-Co-Net and Co-Net for unaware and JPEG-aware scenarios over StyleGAN2 and VIPPrint datasets.}
		\label{tab0}
\begin{tabular}{|c|c|c|c|l}
\cline{1-4}
\textbf{Scenario}                             & \textbf{Network}     & \textbf{StyleGAN2 dataset} & \textbf{VIPPrint dataset} &  \\ \cline{1-4}
\multirow{2}{*}{Unaware Scenario}    & Cross-Co-Net & 99.80\%           & 99.53\%          &  \\ \cline{2-4}
                                     & Co-Net      & 98.25\%           & 98.60\%          &  \\ \cline{1-4}
\multirow{2}{*}{JPEG-aware Scenario} & Cross-Co-Net & 94.40\%           & 93.05\%          &  \\ \cline{2-4}
                                     & Co-Net      & 93.80\%           & 92.56\%          &  \\ \cline{1-4}
\end{tabular}
\end{table}

\begin{figure}[!h]
	\centering
	\subfloat[Red]{{\includegraphics[width=4cm]{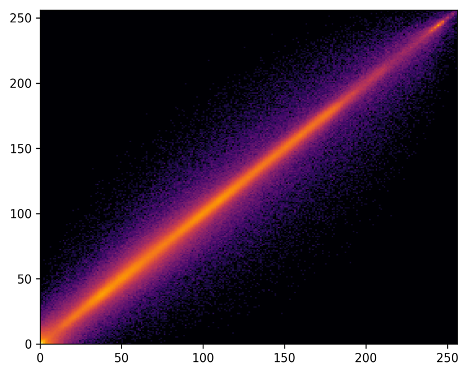} }}%
	\subfloat[Green]{{\includegraphics[width=4cm]{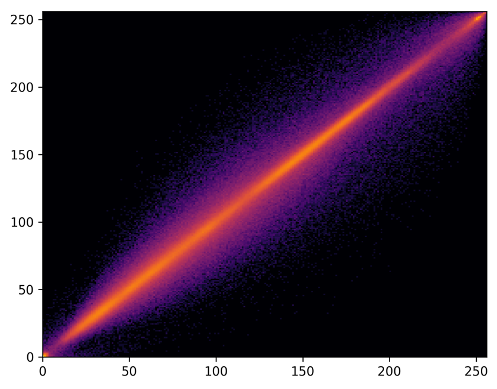} }}%
	\subfloat[Blue]{{\includegraphics[width=4cm]{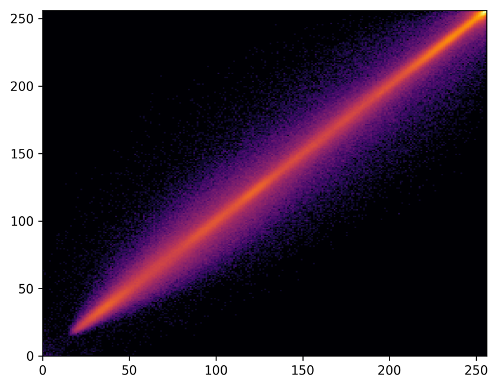} }}%
	\qquad
	\subfloat[Red and Green]{{\includegraphics[width=4cm]{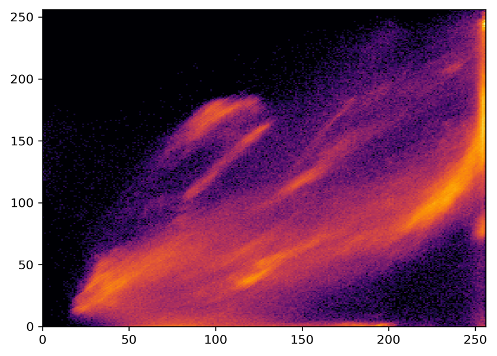} }}%
	\subfloat[Red and Blue]{{\includegraphics[width=4cm]{Figures/RB_RAW.png} }}%
	\subfloat[Green and Blue]{{\includegraphics[width=4cm]{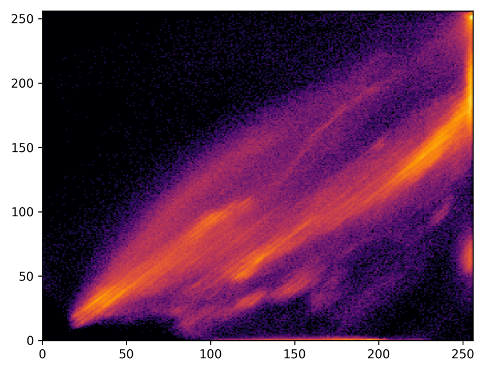} }}%
	\caption{\blue{Visual representation of the Co-occurrence matrices for real image with different channel combinations}}%
	\label{fig:example55}%
\end{figure}

\begin{figure}[!h]%
	\centering
	\subfloat[Red]{{\includegraphics[width=4cm]{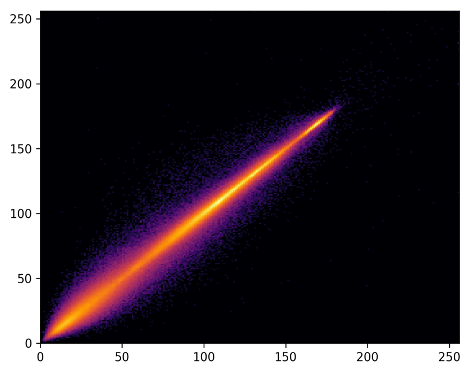} }}%
	\subfloat[Green]{{\includegraphics[width=4cm]{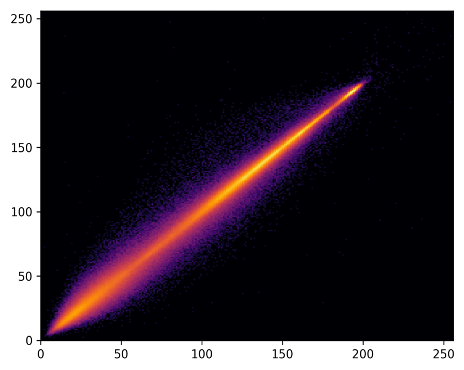} }}%
	\subfloat[Blue]{{\includegraphics[width=4cm]{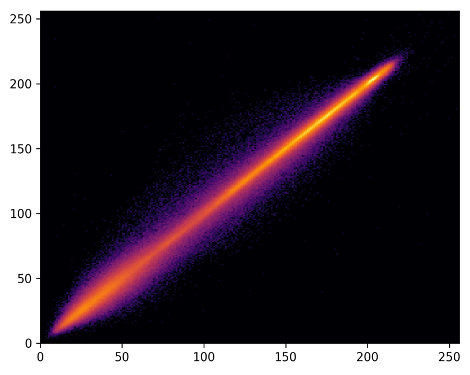} }}%
	\qquad
	\subfloat[Red and Green]{{\includegraphics[width=4cm]{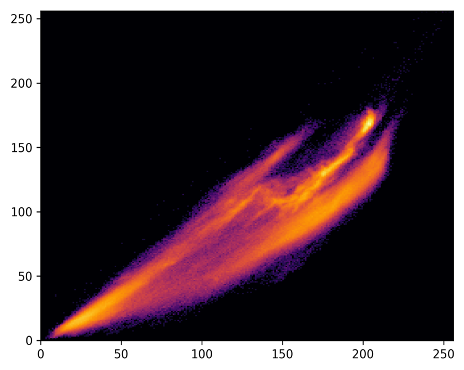} }}%
	\subfloat[Red and Blue]{{\includegraphics[width=4cm]{Figures/RB_GAN.png} }}%
	\subfloat[Green and Blue]{{\includegraphics[width=4cm]{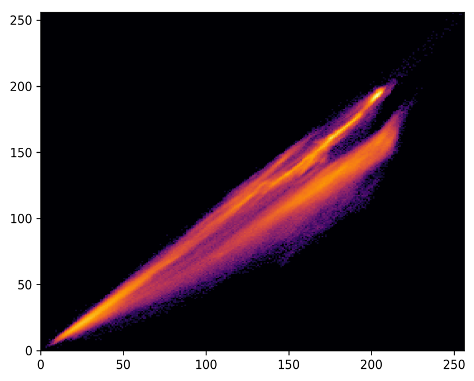} }}%
	\caption{\blue{Visual representation of the Co-occurrence matrices for GAN image StyleGAN2 with different channel combinations}}%
	\label{fig:example66}%
\end{figure}

Furthermore, we illustrate the differences among spectral and spatial bands in a single real image separately in Figure \ref{fig:example55}. In Figure \ref{fig:example66}, we show the discrepancies in a single GAN StyleGAN2 image.

\subsubsection{Cross-Co-Net and Co-Net Robustness.}

We present the accuracies of the tests performed in different post-processing operations in Table \ref{tab3} by considering StyleGAN2 and VIPPrint datasets. We observe that Cross-Co-Net achieves substantially more heightened robustness in all cases, even if the image post-processing is strong. We also remark that the worst-case situation refers to AHE and blurring followed by sharpening, where the Cross-Co-Net accuracy decreases to 75 percent or slightly less. According to the Co-Net's results, we can notice that the network's accuracy is close or equal to 50 percent in many scenarios. The rationale for such a result is that nearly all GAN images are labeled as real after post-processing, implying that the artifacts on which the model turns to detect the GAN-image are removed through post-processing. On the other hand, the Cross-Co-Net performs well despite a small loss in performance, demonstrating that by analyzing cross-band co-occurrences, the network may learn in-depth features the characteristics of GAN images, resulting in a robust model subsequent processing. Furthermore, most processing operations modify the spatial relations across pixels but not the intra-channel relations, providing another reason for Cross-Co-Net's robustness versus post-processing.

\begin{table}[h!]
\caption{Robustness performances in the presence of post-processing for StyleGAN2 and VIPPrint dataset. }
	\label{tab3}
\begin{tabular}{cc|cc|cc|l}
\cline{3-6}
                                                                                                 &                     & \multicolumn{2}{c|}{\textbf{StyleGAN2 dataset}}                           & \multicolumn{2}{c|}{\textbf{VIPPrint dataset}}               &  \\ \cline{1-6}
\multicolumn{1}{|c|}{\textbf{Operations}}                                                        & \textbf{Parameters} & \multicolumn{1}{c|}{\textbf{Cross-Co-Net}} & \textbf{Co-Net}              & \multicolumn{1}{c|}{\textbf{Cross-Co-Net}} & \textbf{Co-Net} &  \\ \cline{1-6}
\multicolumn{1}{|c|}{\multirow{2}{*}{Median filter}}                                             & 3 x 3               & \multicolumn{1}{c|}{96.25\%}               & 50.00\%                      & \multicolumn{1}{c|}{95.15\%}               & 50.00\%         &  \\ \cline{2-6}
\multicolumn{1}{|c|}{}                                                                           & 5 x 5               & \multicolumn{1}{c|}{90.35\%}               & 50.00\%                      & \multicolumn{1}{c|}{89.25\%}               & 50.00\%         &  \\ \cline{1-6}
\multicolumn{1}{|c|}{\multirow{3}{*}{Gaussian noise}}                                            & 0.5                 & \multicolumn{1}{c|}{99.95\%}               & 86.40\%                      & \multicolumn{1}{c|}{97.75\%}               & 83.10\%         &  \\ \cline{2-6}
\multicolumn{1}{|c|}{}                                                                           & 0.8                 & \multicolumn{1}{c|}{99.55\%}               & \multicolumn{1}{l|}{64.10\%} & \multicolumn{1}{c|}{97.87\%}               & 61.20\%         &  \\ \cline{2-6}
\multicolumn{1}{|c|}{}                                                                           & 2                   & \multicolumn{1}{c|}{90.70\%}               & \multicolumn{1}{l|}{50.00\%} & \multicolumn{1}{c|}{87.60\%}               & 50.00\%         &  \\ \cline{1-6}
\multicolumn{1}{|c|}{AHE}                                                                        & -                   & \multicolumn{1}{c|}{75.00\%}               & \multicolumn{1}{l|}{50.00\%} & \multicolumn{1}{c|}{72.00\%}               & 50.00\%         &  \\ \cline{1-6}
\multicolumn{1}{|c|}{\multirow{3}{*}{Gamma correction}}                                          & 0.9                 & \multicolumn{1}{c|}{99.65\%}               & \multicolumn{1}{l|}{55.90\%} & \multicolumn{1}{c|}{97.60\%}               & 52.85\%         &  \\ \cline{2-6}
\multicolumn{1}{|c|}{}                                                                           & 0.8                 & \multicolumn{1}{c|}{82.50\%}               & 50.80\%                      & \multicolumn{1}{c|}{82.30\%}               & 50.00\%         &  \\ \cline{2-6}
\multicolumn{1}{|c|}{}                                                                           & 1.2                 & \multicolumn{1}{c|}{91.70\%}               & 50.15\%                      & \multicolumn{1}{c|}{90.63\%}               & 50.05\%         &  \\ \cline{1-6}
\multicolumn{1}{|c|}{\multirow{2}{*}{Average blurring}}                                          & 3 x 3               & \multicolumn{1}{c|}{92.85\%}               & 72.50\%                      & \multicolumn{1}{c|}{90.80\%}               & 67.00\%         &  \\ \cline{2-6}
\multicolumn{1}{|c|}{}                                                                           & 5 x 5               & \multicolumn{1}{c|}{85.30\%}               & 54.10\%                      & \multicolumn{1}{c|}{81.00\%}               & 50.10\%         &  \\ \cline{1-6}
\multicolumn{1}{|c|}{\multirow{3}{*}{Resizing}}                                                  & 0.9                 & \multicolumn{1}{c|}{99.73\%}               & 90.78\%                      & \multicolumn{1}{c|}{97.70\%}               & 85.91\%         &  \\ \cline{2-6}
\multicolumn{1}{|c|}{}                                                                           & 0.8                 & \multicolumn{1}{c|}{99.50\%}               & 76.65\%                      & \multicolumn{1}{c|}{98.50\%}               & 73.60\%         &  \\ \cline{2-6}
\multicolumn{1}{|c|}{}                                                                           & 0.5                 & \multicolumn{1}{c|}{81.50\%}               & 50.05\%                      & \multicolumn{1}{c|}{80.00\%}               & 50.00\%         &  \\ \cline{1-6}
\multicolumn{1}{|c|}{\multirow{3}{*}{Zooming}}                                                   & 1.1                 & \multicolumn{1}{c|}{99.60\%}               & 94.95\%                      & \multicolumn{1}{c|}{99.00\%}               & 90.05\%         &  \\ \cline{2-6}
\multicolumn{1}{|c|}{}                                                                           & 1.2                 & \multicolumn{1}{c|}{99.45\%}               & 89.95\%                      & \multicolumn{1}{c|}{97.10\%}               & 84.00\%         &  \\ \cline{2-6}
\multicolumn{1}{|c|}{}                                                                           & 1.9                 & \multicolumn{1}{c|}{98.60\%}               & 57.65\%                      & \multicolumn{1}{c|}{96.00\%}               & 52.30\%         &  \\ \cline{1-6}
\multicolumn{1}{|c|}{\multirow{3}{*}{Rotation}}                                                  & 5                   & \multicolumn{1}{c|}{99.45\%}               & 93.65\%                      & \multicolumn{1}{c|}{99.00\%}               & 90.15\%         &  \\ \cline{2-6}
\multicolumn{1}{|c|}{}                                                                           & 10                  & \multicolumn{1}{c|}{99.50\%}               & 93.65\%                      & \multicolumn{1}{c|}{98.50\%}               & 91.60\%         &  \\ \cline{2-6}
\multicolumn{1}{|c|}{}                                                                           & 45                  & \multicolumn{1}{c|}{99.50\%}               & 71.90\%                      & \multicolumn{1}{c|}{98.50\%}               & 70.09\%         &  \\ \cline{1-6}
\multicolumn{1}{|c|}{Cropping}                                                                   & -                   & \multicolumn{1}{c|}{99.80\%}               & 92.60\%                      & \multicolumn{1}{c|}{97.60\%}               & 91.32\%         &  \\ \cline{1-6}
\multicolumn{1}{|c|}{\begin{tabular}[c]{@{}c@{}}Blurring followed \\ by sharpening\end{tabular}} & -                   & \multicolumn{1}{c|}{73.60\%}               & 50.00\%                      & \multicolumn{1}{c|}{73.00\%}               & 51.10\%         &  \\ \cline{1-6}
\end{tabular}
\end{table}

\subsection{Performance and Robustness of JPEG-aware Cross-Co-Net}
In this subsection, we present the effectiveness and robustness of JPEG-aware Cross-Co-Net in the match and mismatch JPEG quality factors.

\subsubsection{Performance under different quality factors.}

As an additional study, we verified that the efficiency of both Cross-Co-Net and Co-Net degrades when JPEG compression is used. Particularly, with QF = [95], detection performance is already less than 90 percent, falling under 80 percent when QF = [85]. Nevertheless, the loss of efficiency using JPEG compression is not unexpected. The majority of authentic images in the FFHQ dataset have been at least once JPEG compressed and thus producing compression footprints. On the other hand, the GAN images do not show such traces. The network may then implicitly associate compression artifacts with the class of actual images. According to our results, the JPEG-aware Cross-Co-Net achieves an average accuracy of 94.40 percent on a test set of JPEG authentic and GAN images when the quality factors [75, 80, 85, 90, and 95] are used for training. Therefore, the model trained on the stated QFs values can be generalized to other compression settings. Table \ref{tabJPEG_StyleGan2} presents the outcomes of tests run on a StyleGAN2 dataset containing both match and mismatch QF values. When mismatch QFs are taken into account, the actual loss of accuracy in the ranges [73, 77, 83, 87, 93, and 97] is less than one percent, suggesting that the network's generalization capabilities are rather impressive. Table \ref{tabJPEG_StyleGan2} also shows the average loss of accuracy for match and mismatch QF values in the VIPPrint dataset. While VIPPrint dataset detection is significantly more complicated in this case, the average accuracy loss is less than three percent.

\begin{table}[!h]
\centering
\caption{Accuracies of the JPEG-aware Cross-Co-Net for matched and mismatched quality factors for StyleGAN2 and VIPPrint dataset.}
	\label{tabJPEG_StyleGan2}
\begin{tabular}{c|cc|l}
\cline{2-3}
\textbf{}                                      & \multicolumn{2}{c|}{\textbf{Accuracy}}                                                           &  \\ \cline{1-3}
\multicolumn{1}{|c|}{\textbf{Quality Factors}} & \multicolumn{1}{l|}{\textbf{StyleGAN2 dataset}} & \multicolumn{1}{l|}{\textbf{VIPPrint dataset}} &  \\ \cline{1-3}
\multicolumn{1}{|c|}{73}                       & \multicolumn{1}{c|}{95.40\%}                    & 93.35\%                                        &  \\ \cline{1-3}
\multicolumn{1}{|c|}{75}                       & \multicolumn{1}{c|}{95.83\%}                    & 93.80\%                                        &  \\ \cline{1-3}
\multicolumn{1}{|c|}{77}                       & \multicolumn{1}{c|}{95.93\%}                    & 93.90\%                                        &  \\ \cline{1-3}
\multicolumn{1}{|c|}{80}                       & \multicolumn{1}{c|}{96.35\%}                    & 96.00\%                                        &  \\ \cline{1-3}
\multicolumn{1}{|c|}{83}                       & \multicolumn{1}{c|}{95.73\%}                    & 93.60\%                                        &  \\ \cline{1-3}
\multicolumn{1}{|c|}{85}                       & \multicolumn{1}{c|}{96.28\%}                    & 93.08\%                                        &  \\ \cline{1-3}
\multicolumn{1}{|c|}{87}                       & \multicolumn{1}{c|}{96.50\%}                    & 96.55\%                                        &  \\ \cline{1-3}
\multicolumn{1}{|c|}{90}                       & \multicolumn{1}{c|}{95.70\%}                    & 93.78\%                                        &  \\ \cline{1-3}
\multicolumn{1}{|c|}{93}                       & \multicolumn{1}{c|}{96.60\%}                    & 96.00\%                                        &  \\ \cline{1-3}
\multicolumn{1}{|c|}{95}                       & \multicolumn{1}{c|}{96.10\%}                    & 96.41\%                                        &  \\ \cline{1-3}
\multicolumn{1}{|c|}{97}                       & \multicolumn{1}{c|}{95.80\%}                    & 93.75\%                                        &  \\ \cline{1-3}
\end{tabular}
\end{table}

\subsubsection{Robustness under different quality factors.}

\begin{table}[h!]
\renewcommand\arraystretch{1.3}
	\centering
	\setlength{\tabcolsep}{3pt}
	\caption{JPEG-aware Cross-Co-Net robustness performance with the post-processing operators (StyleGAN2). }
	\label{tabJPEG-PP-StyleGan2}
	\scalebox{1}{
	\begin{tabular}{|c|c|c|c|c|c|}	
		\hline
\textbf{QFs}  & \textbf{Median Filtering} & \textbf{Resizing} & \textbf{Gaussian Noise} & \textbf{Zooming} & \textbf{AHE} \\ \hline
73 & 93.80\% & 92.10\% & 93.00\% & 92.00\% & 91.00\% \\ \hline
75 & 94.10\% & 93.10\% & 92.80\% & 94.80\% & 92.70\% \\ \hline
77 & 94.20\% & 93.70\% & 93.30\% & 94.30\% & 92.20\% \\ \hline
80 & 94.00\%&  94.40\% & 93.50\% & 94.50\% & 93.60\% \\ \hline
83 & 94.00\% &  94.00\%& 93.01\% & 94.50\% & 92.05\%		\\ \hline
85 & 93.70\% & 94.10\%& 93.40\% & 93.40\% & 92.00\% \\ \hline
87 & 94.10\% & 94.40\%& 93.40\% & 94.40\% & 92.00\%\\ \hline
90 & 94.10\% & 94.40\% & 93.10\%  & 94.10\% & 92.10\%\\ \hline
93 & 94.10\%& 94.30\% & 91.10\% & 94.10\%  & 90.10\% \\ \hline
95 & 94.10\% & 94.80\%& 88.30\%  & 94.30\%  & 86.04\% \\ \hline
97 & 94.40\% & 94.60\%& 88.05\%  & 94.06\% & 86.00\% \\ \hline
	\end{tabular}
	}
\end{table}
\begin{table}[h!]
\renewcommand\arraystretch{1.3}
	\centering
	\setlength{\tabcolsep}{3pt}
	\caption{JPEG-aware Cross-Co-Net robustness performance with the post-processing operators (VIPPrint).}
	\label{tabJPEG-PP-VIPPrint}
	\scalebox{1}{
	\begin{tabular}{|c|c|c|c|c|c|}	
		\hline
\textbf{QFs}  & \textbf{Median Filtering} & \textbf{Resizing} & \textbf{Gaussian Noise} & \textbf{Zooming} & \textbf{AHE} \\ \hline
73 & 92.30\% & 90.60\% & 90.09\% & 90.80\% & 90.25\% \\ \hline
75 & 93.00\% & 91.70\% & 90.50\% & 92.72\% & 91.00\% \\ \hline
77 & 93.14\% & 91.40\% & 91.10\% & 92.37\% & 91.10\% \\ \hline
80 & 93.05\%&  92.70\% & 91.70\% & 92.57\% & 92.50\% \\ \hline
83 & 93.18\% &  92.20\%& 91.00\% & 92.58\% & 91.55\%		\\ \hline
85 & 92.40\% & 92.00\%& 91.30\% & 91.44\% & 91.00\% \\ \hline
87 & 93.00\% & 92.20\%& 91.30\% & 92.44\% & 91.00\%\\ \hline
90 & 93.00\% & 92.80\% & 91.00\%  & 92.20\% & 91.02\%\\ \hline
93 & 93.00\%& 92.70\% & 90.00\% & 92.17\%  & 87.00\% \\ \hline
95 & 93.00\% & 92.50\%& 83.20\%  & 92.33\%  & 84.44\% \\ \hline
97 & 93.21\% & 92.50\%& 83.15\%  & 92.00\% & 84.05\% \\ \hline
	\end{tabular}
	}
\end{table}

We proved that the efficiency generated by the JPEG-aware Cross-Co-Net network is quite small when processing operations are performed before the final compression. For the scenarios of median filtering with a window size of $5\times 5$, resizing with a scaling factor of 0.8, Gaussian noise with a standard deviation of 2, Zooming with a scaling factor of 1.9, and AHE, we report the results in Tables \ref{tabJPEG-PP-StyleGan2}, and \ref{tabJPEG-PP-VIPPrint}, by considering StyleGAN2, and VIPPrint, respectively. This study reveals that the JPEG-aware edition of Cross-Co-Net maintains the model's sufficient robustness, indicating that concentrating on color band discrepancies via cross-band co-occurrences is an efficient method to discriminate between authentic and GAN-generated images.

\section{Conclusion and Future Works}
In this chapter, we proposed a CNN method for detecting high-quality GAN-generated images, emphasizing fake face identification. Our suggested approach makes use of spectral band discrepancies and pixel co-occurrence matrices. Furthermore, we used cross-band co-occurrence matrices to construct the CNN architecture for extracting discriminant features for the authentic and GAN classes. The experimental results demonstrate the performance of the proposed approach. Finally, we demonstrated the resilience of our approach to post-processing comparing to single spatial co-occurrences to train the detector, thus illustrating the relevance of considering color band correlations. \\
In the upcoming work, research effort will be devoted to Cross-Co-Net's performance in both a white-box and a black-box scenario when the band relationships are purposefully changed to mislead the detectors. In this case, it is important to assess the network's effectiveness against a knowledgeable adversary who performs adversarial examples against the CNN model to estimate the co-occurrence computation and backpropagate the gradients to the pixel domain.

\bibliographystyle{splncs04}
\bibliography{Main}

\end{document}